\documentclass[10pt]{article}
\usepackage{latexsym}
\usepackage{amssymb}
\usepackage{amsmath}
\usepackage{amscd}
\usepackage{amsthm}
\usepackage[left=2cm,top=2.5cm,right=2.5cm,bottom=1.5cm]{geometry}
\usepackage[dvips]{graphicx}
\usepackage{hyperref}
\begin{document}
\begin{center}
\large{\bf{A New Class of Bianchi Type-I Cosmological Models in Scalar-Tensor Theory of Gravitation and Late Time Acceleration}} \\
\vspace{10mm}
\normalsize{Anirudh Pradhan$^1$, Ajay Kumar Singh$^2$, H. Amirhashchi$^3$}\\
\vspace{5mm}
\normalsize{$^{1,2}$Department of Mathematics, Hindu Post-graduate College, Zamania-232 331, Ghazipur, India \\
\vspace{2mm}
$^1$e-mail: pradhan@iucaa.ernet.in; pradhan.anirudh@gmail.com }\\
\vspace{5mm}
\normalsize{$^3$Young Researchers Club, Mahshahr Branch, Islamic Azad University, Mahshahr, Iran \\
\vspace{2mm}
e-mail: h.amirhashchi@mahshahriau.ac.ir} \\
\end{center}
\vspace{10mm}
%\date{}
%\maketitle
\begin{abstract} 
A new class of a spatially homogeneous and anisotropic Bianchi type-I cosmological models of the universe 
for perfect fluid distribution within the framework of scalar-tensor theory of gravitation proposed by 
S$\acute{a}$ez and Ballester (Phys. Lett. 113:467, 1986) is investigated. To prevail the deterministic solutions 
we choose the different scale factors which yield time-dependent deceleration parameters (DP) representing 
models which generate a transition of the universe from the early decelerated phase to the recent accelerating 
phase. Three different physically viable models of the universe are obtained in which their anisotropic solutions 
may enter to some isotropic inflationary era. The modified Einstein's field equations are solved exactly 
and the models are found to be in good concordance with recent observations. Some physical and geometric 
properties of the models are also discussed. 
\end{abstract}
%\newpage
%%%%%%%%%%%%%%%%%%%%%%%%%%%%%%%%%%%%%%%%%%%%%%%%%%%%%%%%%%%%%%%%%%%%%%%%%%%%%%%%%%%%%%%%%%%%%%%%%%%%%
%%%%%%%%%%%%%%%%%%%%%%%%%%%%%%%%%%%% SECTION 1  %%%%%%%%%%%%%%%%%%%%%%%%%%%%%%%%%%%%%%%%%%%%%%%%%%%%%
\section{Introduction} 
General relativity (GR) passes all present tests with flying colours, however, there are several reasons why 
it remains very important to consider alternative theories of gravity. The first one is that theoretical 
attempts at quantizing gravity or unifying it with other interactions generically predict deviation from 
Einstein's theory, because gravitation is no longer mediated by a pure spin-2 field but also by partners 
to the graviton. The second reason is that it is any way extremely instructive to contrast GR's predictions 
with those of alternative models, even if there were no serious theoretical motivation for them. The third 
reason is the existence of several puzzling experimental issues, which do not contradict GR in a direct way, 
but may nevertheless suggest that gravity does not behave at large distances exactly as Newton and Einstein 
predicted. Cosmological observations notably tell us that about $96\%$ of the total energy density of the 
universe is composed of unknown, non-baryonic, fluids ($72\%$ of ``dark energy'' and $24\%$ of ``dark matter''), 
and the acceleration of the two Pioneer spacecrafts towards the Sun happens to be larger than what is expected 
from the $1/r^{2}$ law. Therefore, alternative theories are any way important to study. \\ 

Since the observed universe is almost homogeneous and isotropic, space-time is usually described by a 
Friedman-Lemaitre-Robertson-Walker (FLRW) cosmology. But it is also believed that in the early universe the 
FLRW model does not give a correct matter description. The anomalies found in the cosmic microwave background 
(CMB) and the large structure observations stimulated a growing interest in anisotropic cosmological model 
of the universe. Observations by the Differential Macrowave Radiometers (DMR) on NASA's Cosmic Background Explorer 
Spacecraft registered anisotropy in various angle scales. It is conjectured, that these anisotropies hide in 
their hearts the entire history of the cosmic evolution down to recombination, and they are considered to be 
indicative of the universe geometry and the matter composing the universe. It is expected, that much more will 
be known about anisotropy of cosmic microwave's background after the investigations of the microwave's anisotropy 
probe. There is a general agreement among cosmologists that cosmic microwave's background anisotropy in the small 
angle scale holds the key to the formation of the discrete structure. The theoretical argument \cite{ref1} and 
the modern experimental data support the existence of an anisotropic phase, which turns into an isotropic one.  \\

In recent years, there has been considerable interest in scalar tensor theories of gravitation which 
are considered to be essential to describe the gravitational interaction near the planks scale, string 
theory, extended inflation and many higher order theories imply scalar field. Scalar-Tensor theories of 
gravitation provide the natural generalizations of general relativity and also provide a convenient set 
of representations for the observational limits on possible deviations from general relativity. There are 
two categories of gravitational theories involving a classical scalar field $\phi$. In first category the 
scalar field $\phi$ has the dimension of the inverse of the gravitational constant $\rm G$ among which the 
Brans-Decke theory \cite{ref2} is of considerable importance and the role of the scalar field is 
confined to its effect on gravitational field equations. Brans and Decke formulated a scalar-tensor theory 
of gravitation which introduces an additional scalar field $\phi$ besides the metric tensor $g_{ij}$ and a 
dimensionless coupling constant $\omega$. This theory goes to general relativity for large values of the 
coupling constant $\omega > 500$. In the second category of theories involve a dimensionless scalar field. 
S$\acute{a}$ez and Ballester \cite{ref3} developed a scalar-tensor theory in which the metric is coupled with a 
dimensionless scalar field in a simple manner. This coupling gives a satisfactory description of the weak 
fields. In spite of the dimensionless character of the scalar field, an anti-gravity regime appears. This 
theory suggests a possible way to solve the missing-matter problem in non-flat FRW cosmologies. The 
Scalar-Tensor theories of gravitation play an important role to remove the graceful exit problem in the 
inflation era \cite{ref4}. In earlier literature, cosmological models within 
the framework of S$\acute{a}$ez-Ballester scalar-tensor theory of gravitation, have been studied by Singh and 
Agrawal \cite{ref5,ref6}, Ram and Tiwari \cite{ref7}, Singh and Ram \cite{ref8}. Mohanty and Sahu \cite{ref9,ref10} 
have studied Bianchi type-$\rm VI_{0}$ and Bianchi type-I models in Saez-Ballester theory. In recent years, Tripathi et al. 
\cite{ref11}, Reddy et al. \cite{ref12,ref13}, Reddy and Naidu \cite{ref14}, Rao et al. \cite{ref15,ref16,ref17}, Adhav et al. 
\cite{ref18}, Katore et al. \cite{ref19}, Sahu \cite{ref20}, Singh \cite{ref21} and Pradhan and Singh \cite{ref22} 
have obtained the solutions in S$\acute{a}$ez-Ballester scalar-tensor theory of gravitation in different context. 
Recently, Naidu et al. \cite{ref23} and Reddy et al. \cite{ref24} have studied LRS Bianchi type-II and five dimensional 
dark energy models in S$\acute{a}$ez and Ballester scalar tensor theory of gravitation respectively. \\

Recently, Kumar and Singh \cite{ref25} obtained exact Bianchi type-I cosmological models in S$\acute{a}$ez and Ballester 
Scalar-Tensor theory of gravitation by assuming the constant deceleration parameter. In literature it is common 
to use a constant deceleration parameter, as it duly gives a power law for metric function or corresponding quantity. 
Several cosmological observations indicate that the observable universe is undergoing a phase of accelerated expansion 
(Riess et al. \cite{ref26}; Perlmutter et al. \cite{ref27}; Bennet et al. \cite{ref28}; Tegmark et al. \cite{ref29}; 
Allen et al. \cite{ref30}). Recent observations of SNe Ia of high confidence level (Tonry et al. \cite{ref31}; Riess et 
al. \cite{ref32}; Clocchiatti et al. \cite{ref33}) have further confirmed this. Also, the transition redshift from 
deceleration expansion to accelerated expansion is about 0.5. Now for a Universe which was decelerating in past and 
accelerating at the present time, the DP must show signature flipping (see the Refs. Padmanabhan and Roychowdhury 
\cite{ref34}, Amendola \cite{ref35}, Riess et al. \cite{ref36}). So, in general, the DP is not a constant but time 
variable. \\

Socorro et al. \cite{ref37} and Jamil et al. \cite{ref38} have studied on anisotropic Bianchi type-I cosmology in 
S$\acute{a}$ez-Ballester scalar-tensor theory of gravity. Recently, Pradhan et al. \cite{ref39} have studied some 
exact Bianchi type-I cosmological models in scalar-tensor theory of gravitation with time dependent deceleration 
parameter. Motivated by the discussions, in this paper, we have obtained a new class of exact solutions of field 
equations given by S$\acute{a}$ez and Ballester \cite{ref3} in Bianchi type-I space-time by considering a time 
dependent deceleration parameter. The out line of the paper is as follows: In Section $2$, the metric with a 
Lagrangian and the field equations are described. Section $3$ deals with the solutions of the field equations. 
Subsections $3.1$, $3.2$ and $3.3$ describe the three different models by considering the three different values 
of scale factors which yield the time dependent deceleration parameters and physical and geometric behaviour of the 
model are also discussed. Finally, conclusions are summarized in the last Section $4$. \\\\ 
%%%%%%%%%%%%%%%% %%%%%%%%%%%%%%%%%%%%%%%%%%%%%%%%%%%%%%%%%%%%%%%%%%%%%%%%%%%%%%%%%%%%%%%%%%%%%%%%%%%
%%%%%%%%%%%%%%%%%%%%%%%%%%%%%%%  SECTION 2  %%%%%%%%%%%%%%%%%%%%%%%%%%%%%%%%%%%%%%%%%%%%%%%%%%%%%%%
\section{The Metric and Field Equations}
We consider spatially homogeneous and anisotropic Bianchi type-$I$ space-time given by
\begin{equation}
\label{eq1}
ds^{2} = - dt^{2}+ A^{2}dx^{2} + B^{2} dy^{2} + C^{2} dz^{2},
\end{equation}
where the metric potentials $A$, $B$ and $C$ are functions of cosmic time $t$ alone. This ensures that the model is
spatially homogeneous. \\

We, consider the simple case of a homogeneous but anisotropic Bianchi type-I model with matter term with a scalar 
field $\phi$. Our model is based on a non-standard scalar-tensor theory, defined in S$\acute{a}$ez and Ballester 
\cite{ref3} with a dimensionless scalar field $\phi$ and tensor field $g_{ij}$. This alternative theory of gravitation 
is combined scalar and tensor fields in which the metric is coupled with a dimensionless scalar field. We assume the Lagrangian
\begin{equation}
\label{eq2} L = R - \omega \phi^{k} \phi_{,i}\phi^{,i},
\end{equation}
$R$ being the scalar curvature, $\phi$ a dimensionless scalar field, $\omega$ and $k$ arbitrary dimensionless constants and 
$\phi^{,i}$ the contraction $\phi_{,\alpha}g^{\alpha i}$. Here the partial derivatives are denoted by a comma and the covariant 
derivatives by a semicolon in the usual way. \\

From the above Lagrangian we can establish the action
\begin{equation}
\label{eq3} I = \int_{\sum}{\left(L + \chi L_{m}\right)(-g)^{\frac{1}{2}}}dx^{1}dx^{2}dx^{3}dx^{4},
\end{equation}
where $L_{m}$ is the matter Lagrangian, $g$ is the determinant of the matrix $g_{ij}$, $x^{i}$ are the coordinates, $\sum$ 
is an arbitrary region of integration and $\chi = 8\pi$ (we use geometrized units). When $k = 0$, our model is just 
the Einstein gravity with a massless minimally coupled scalar field coupled to gravity. Different aspects of this model 
have been investigated in the literatures (already described in Sect. $1$). By considering arbitrary independent variations 
of the metric and the scalar field vanishing at the boundary of $\sum$, the variation principle
\begin{equation}
\label{eq4} \delta{I} = 0,
\end{equation}
leads to a generalized Einstein equation 
\[
G_{ij} - \omega \phi^{k}\left(\phi_{,i} \phi_{,j} - \frac{1}{2}g_{ij}\phi_{,l}\phi^{,l}\right) = - 8\pi T_{ij},
\]
\begin{equation}
\label{eq5} 2\phi^{k}\phi_{;i}^{,i} + k \phi^{k -1}\phi_{,l}\phi^{,l} = 0,
\end{equation}
where $G_{ij} = R_{ij} - \frac{1}{2}R g_{ij}$ is the Einstein tensor; $T_{ij}$ is the stress-energy tensor of the matter
Lagrangian $L_{m}$.  \\

Since the action $I$ is a scalar, it can be easily proved that the equation of motion
\begin{equation}
\label{eq6} T^{ij}_{~ ;i} = 0,
\end{equation}
are consequences of the field equations. \\

In Brans-Dicke gravity, the modification was introduced due to the lack of compatibility of Einstein's theory with 
Mach's principle. But in S$\acute{a}$ez and Ballester \cite{ref3} a scalar-tensor theory of gravity was introduced in which 
metric is coupled to scalar field. Here the strength of the coupling between gravity and the field was governed by a 
parameter $\omega$. With this modification, they were able to solve a `missing-mass problem'. Several aspects of 
S$\acute{a}$ez \& Ballester scalar-tensor theory of gravitation in relation to Bianchi cosmological models have been 
explored in literature.\\

The energy-momentum tensor $T_{ij}$ for a perfect fluid has the form
\begin{equation}
\label{eq7} T_{ij} = (p + \rho)u_{i}u_{j} - p g_{ij},
\end{equation}
where $p$ is the thermodynamical pressure, $\rho$ the energy density, $u_{i}$ the four-velocity of the fluid  satisfying
\begin{equation}
\label{eq8} g_{ij}u^{i}u^{j} = 1.
\end{equation}
In co-moving system of coordinates, we have $u_{i} = (0, 0, 0, 1)$. For the energy momentum tensor (\ref{eq7}) and 
Bianchi type-$I$ space-time (\ref{eq1}), Einstein's modified field equations (\ref{eq5}) yield the 
following five independent equations
\begin{equation}
\label{eq9} \frac{\ddot{A}}{A} + \frac{\ddot{B}}{B} + \frac{\dot{A}\dot{B}}{AB}  = - 8\pi p + \frac{1}{2}\omega 
\phi^{k}\dot{\phi}^{2},
\end{equation}
\begin{equation}
\label{eq10} \frac{\ddot{A}}{A} + \frac{\ddot{C}}{C} + \frac{\dot{A}\dot{C}}{AC} = - 8\pi p + \frac{1}{2}\omega 
\phi^{k}\dot{\phi}^{2},
\end{equation}
\begin{equation}
\label{eq11} \frac{\ddot{B}}{B} + \frac{\ddot{C}}{C} + \frac{\dot{B}\dot{C}}{BC} = - 8\pi p + \frac{1}{2}\omega 
\phi^{k}\dot{\phi}^{2},
\end{equation}
\begin{equation}
\label{eq12} \frac{\dot{A}\dot{B}}{AB} + \frac{\dot{A}\dot{C}}{AC} + \frac{\dot{B}\dot{C}}{BC} = 8\pi \rho - 
\frac{1}{2}\omega \phi^{k}\dot{\phi}^{2},
\end{equation}
\begin{equation}
\label{eq13} \ddot{\phi} + \dot{\phi}\left(\frac{\dot{A}}{A} + \frac{\dot{B}}{B} + \frac{\dot{C}}{C}\right) + 
\frac{m\dot{\phi}^{2}}{2\phi} = 0,
\end{equation}
where an over dot denotes derivative with respect to cosmic time $t$. 
The law of energy-conservation equation ($T^{ij}_{;j} = 0$) gives
\begin{equation}
\label{eq14} \dot{\rho} + (\rho + p) \left(\frac{\dot{A}}{A} + \frac{\dot{B}}{B} + \frac{\dot{C}}{C} \right) = 0.
\end{equation} 
It is worth noting here that our approach suffers from a lack of Lagrangian approach. There is no known way to present 
a consistent
Lagrangian model satisfying the necessary conditions discussed in this paper. \\

The spatial volume for the model (\ref{eq1}) is given by
\begin{equation}
\label{eq15} V^{3} = ABC.
\end{equation}
We define $a = (ABC)^{\frac{1}{3}}$ as the average scale factor so that the Hubble's parameter is anisotropic and may 
be defined as
\begin{equation}
\label{eq16} H = \frac{\dot{a}}{a} = \frac{1}{3}\left(\frac{\dot{A}}{A} + \frac{\dot{B}}{B} + \frac{\dot{C}}{C}\right).
\end{equation}
We also have
\begin{equation}
\label{eq17} H = \frac{1}{3}\left(H_{x} + H_{y} + H_{z}\right),
\end{equation}
where $H_{x} = \frac{\dot{A}}{A}$, $H_{y} = \frac{\dot{B}}{B}$ and $H_{z} = \frac{\dot{C}}{C}$. \\

The deceleration parameter $q$, the scalar expansion $\theta$, shear scalar $\sigma^{2}$ and the average anisotropy 
parameter $A_{m}$ are defined by
\begin{equation}
\label{eq18} q = - \frac{a\ddot{a}}{\dot{a}^{2}} = - \frac{\ddot{a}}{aH^{2}},
\end{equation}
\begin{equation}
\label{eq19}
\theta = \frac{\dot{A}}{A} + \frac{\dot{B}}{B} + \frac{\dot{C}}{C},
\end{equation}
\begin{equation}
\label{eq20}
\sigma^{2} = \frac{1}{2}\left(\sum^{3}_{i = 1}H^{2}_{i} - \frac{1}{3}\theta^{2}\right),
\end{equation}
\begin{equation}
\label{eq21} A_{m} = \frac{1}{3}\sum_{i = 1}^{3}{\left(\frac{\triangle
H_{i}}{H}\right)^{2}},
\end{equation}
where $\triangle H_{i} = H_{i} - H (i = x, y, z)$.

%%%%%%%%%%%%%%%%%%%%%%%%%%%%%%%%%%%%%%%%%%%%%%%%%%%%%%%%%%%%%%%%%%%%%%%%%%%%%%%%%%%%%%%%%%%%%%%%%%
%%%%%%%%%%%%%%%%%%%%%%%%%%%%%%%  SECTION 3  %%%%%%%%%%%%%%%%%%%%%%%%%%%%%%%%%%%%%%%%%%%%%%%%%%%%%%
\section{Solution of the Field Equations}
The field equations (\ref{eq9})-(\ref{eq13}) are five equations necessitating six unknowns $A$, $B$, $C$, $p$, $\rho$ 
and $\phi$. One additional constraint relating these parameters is required to obtain explicit solutions of the system. 
We consider the deceleration parameter as time dependent to get the deterministic solution. \\

We have revisited the solutions of Kumar and Singh \cite{ref25}. Now subtracting (\ref{eq9}) from (\ref{eq10}), (\ref{eq9}) 
from (\ref{eq11}), (\ref{eq10}) from (\ref{eq11}) and taking second integral of each expression, we obtain the following three 
relations, respectively
\begin{equation}
\label{eq22}
\frac{A}{B} = d_{1}\exp{\left(x_{1}\int{a^{-3}dt}\right)},
\end{equation}
\begin{equation}
\label{eq23}
\frac{A}{C} = d_{2}\exp{\left(x_{2}\int{a^{-3}dt}\right)},
\end{equation}
\begin{equation}
\label{eq24}
\frac{B}{C} = d_{3}\exp{\left(x_{3}\int{a^{-3}dt}\right)}, 
\end{equation}
where $d_{1}$, $d_{2}$, $d_{3}$, $x_{1}$, $x_{2}$ and $x_{3}$ are integrating constants. \\

From above Eqs. (\ref{eq22})-(\ref{eq24}), the metric functions $A(t)$, $B(t)$ and $C(t)$ are explicitly obtained as 
\begin{equation}
\label{eq25}
A(t) = a_{1} a \exp{\left(b_{1}\int{a^{-3}dt}\right)},
\end{equation}
\begin{equation}
\label{eq26}
B(t) = a_{2} a \exp{\left(b_{2}\int{a^{-3}dt}\right)},
\end{equation}
\begin{equation}
\label{eq27}
C(t) = a_{3} a \exp{\left(b_{3}\int{a^{-3}dt}\right)},
\end{equation}
where
\[
 a_{1} = (d_{1}d_{2})^{\frac{1}{3}}, \; \; a_{2} = \left(\frac{d_{3}}{d_{1}}\right)^{\frac{1}{3}}, \; \; a_{3} = 
(d_{2}d_{3})^{-\frac{1}{3}},
\]
\begin{equation}
\label{eq28}
b_{1} = \frac{1}{3}(x_{1} + x_{2}), \; \; b_{2} = \frac{1}{3}(x_{3} - x_{1}), \; \; b_{3} 
 = -\frac{1}{3}(x_{2} + x_{3}).
\end{equation}
It is worth mention here that these constants also satisfy the following two conditions
\begin{equation}
\label{eq29}
a_{1}a_{2}a_{3} = 1, \; \; b_{1} + b_{2} + b_{3} = 0.
\end{equation}
The second integral of (\ref{eq13}) leads to
\begin{equation}
\label{eq30}
\phi(t) = \left[\frac{h(k + 2)}{2}\int{a^{-3}dt}\right]^{\frac{2}{(m + 2}},
\end{equation}
where $h$ is a constant due to first integral while the constant of second integral is taken as zero for simplicity 
without any loss of generality. \\
%%%%%%%%%%%%%%%%%%%%%%%%%%%%%%%%%%%%%%%%%%%%%%%%%%%%%%%%%%%%%%%%%%%%%%%%%%%%%%%%%%%%%%%%%%%%%%%%%%
%%%%%%%%%%%%%%%%%%%%%%%%%%%%%%%  SUBSECTION 3  %%%%%%%%%%%%%%%%%%%%%%%%%%%%%%%%%%%%%%%%%%%%%%%%%%%%%%
\subsection{Case 1: when $ a(t) = \sqrt{t^{n}e^{t}}$}

Following Saha et al. \cite{ref40} and Pradhan and Amirhashchi \cite{ref41}, we take following {\it ansatz} for the scale factor, 
where increase in term of time evolution is
\begin{equation}
\label{eq31} a(t) = \sqrt{t^{n}e^{t}},
\end{equation}
where $n$ is a positive constant. This {\it ansatz} generalized the one proposed by Amirhashchi et al. \cite{ref42}. The motivation 
to choose such scale factor (\ref{eq31}) yields a time-dependent DP. Recently Yadav \cite{ref43} has studied such type of 
scale factor in Bianchi-V string cosmological model at late time acceleration. \\

%%%%%%%%%%%%%%%%%%% Figure 1 %%%%%%%%%%%%%%%%%%%%%%%%%%%%%%%%%%%%%%%%%%%%%%%%%%%%%%%%%%%%%
\begin{figure}[ht]
\centering
\includegraphics[width=12cm,height=9.5cm,angle=0]{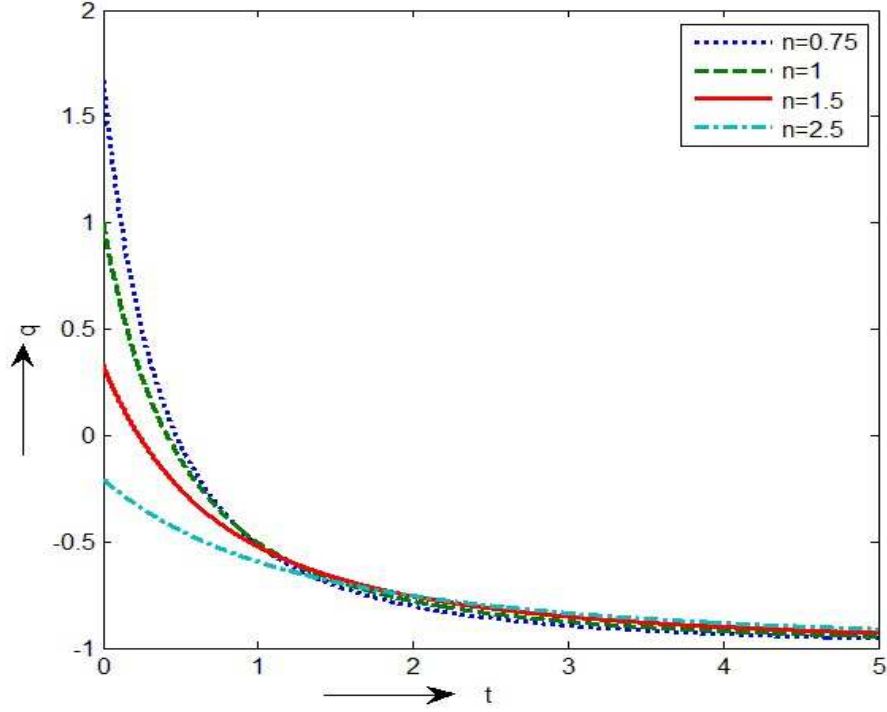} \\
\caption{The plot of deceleration parameter $q$ versus $t$}.
\end{figure}
%%%%%%%%%%%%%%%%%%%%%%%%%%%%%%%%%% %%%%%%%%%%%%%%%%%%%%%%%%%%%%%%%%%%%%%%%%%%%%%%%%%%%%%%%%%%

Substituting Eq. (\ref{eq31}) into Eq. (\ref{eq18}), we find
\begin{equation}
\label{eq32}q = \frac{2n}{(n + t)^{2}} - 1.
\end{equation}
From Eq. (\ref{eq32}), we observe that $q > 0$ for $t < \sqrt{2n} - n$ and $q < 0$ for $t > \sqrt{2n} -n$. 
It is observed that for $0 < n < 2$, our model is evolving from deceleration phase to acceleration phase. Also, recent 
observations of SNe Ia, expose that the present universe is accelerating and the value of DP lies to some place in the 
range $-1 < q < 0$. It follows that in our derived model, one can choose the value of DP consistent with the observation. 
Figure $1$ graphs the deceleration parameter ($q$) versus time which gives the behaviour of $q$ from decelerating to 
accelerating phase for different values of $n$ which is consistent with recent observations of Type Ia supernovae 
(Riess et al. \cite{ref26}; Perlmutter et al. \cite{ref27}; Tonry et al. \cite{ref31}; Riess et al. \cite{ref32}; 
Clocchiatti et al. \cite{ref33}). \\

Using Eq. (\ref{eq31}) in (\ref{eq25})-(\ref{eq27}), we get the following expressions for metric coefficients
\begin{equation}
\label{eq33}
A = a_{1} (\sqrt{t^{n}e^{t}}) \exp{\left[b_{1}\int{(t^{n}e^{t})^{-\frac{3}{2}}}dt\right]},
\end{equation}
\begin{equation}
\label{eq34}
B = a_{2} (\sqrt{t^{n}e^{t}})\exp{\left[b_{2}\int{(t^{n}e^{t})^{-\frac{3}{2}}}dt\right]},
\end{equation}
\begin{equation}
\label{eq35}
C = a_{3} (\sqrt{t^{n}e^{t}})\exp{\left[b_{3}\int{(t^{n}e^{t})^{-\frac{3}{2}}}dt\right]}.
\end{equation}
%%%%%%%%%%%%%%%%%%% Figure 2 %%%%%%%%%%%%%%%%%%%%%%%%%%%%%%%%%%%%%%%%%%%%%%%%%%%%%%%%%%%%%
\begin{figure}[ht]
\centering
\includegraphics[width=10cm,height=10cm,angle=0]{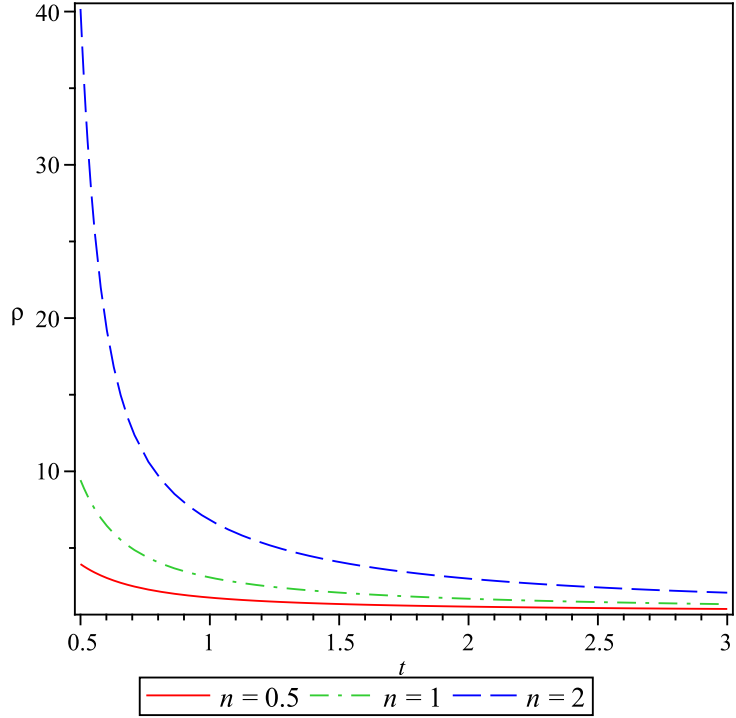} \\
\caption{The plot of energy density $\rho$ versus $t$.
Here $\omega = 1$, $h = \beta_{2} = 1$}.
\end{figure}
%%%%%%%%%%%%%%%%%%%%%%%%%%%%%%%%%% %%%%%%%%%%%%%%%%%%%%%%%%%%%%%%%%%%%%%%%%%%%%%%%%%%%%%%%%%%
%%%%%%%%%%%%%%%%%%% Figure 3 %%%%%%%%%%%%%%%%%%%%%%%%%%%%%%%%%%%%%%%%%%%%%%%%%%%%%%%%%%%%%
\begin{figure}[ht]
\centering
\includegraphics[width=10cm,height=10cm,angle=0]{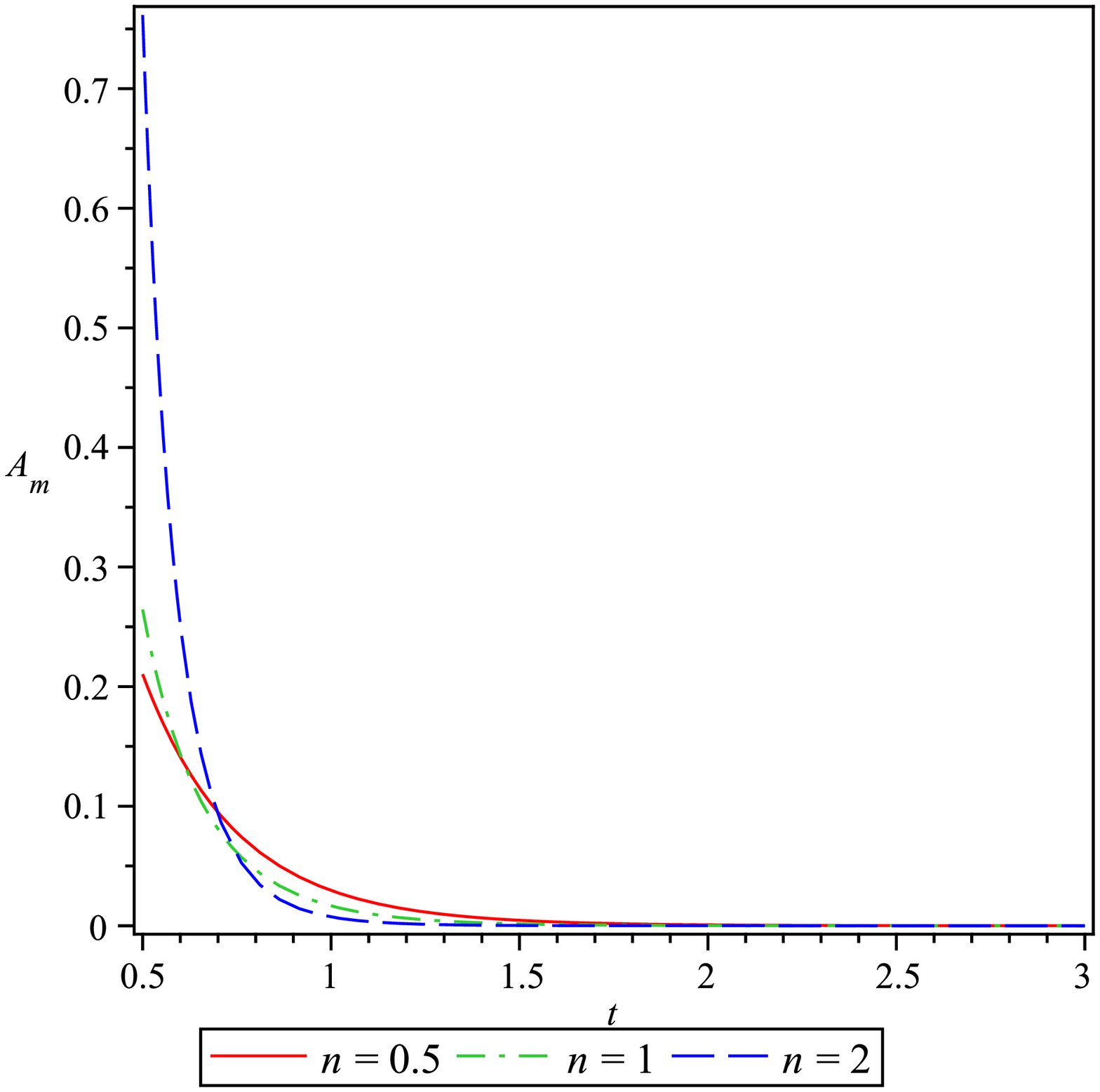} \\
\caption{The plot of anisotropic parameter $A_{m}$ versus $t$. 
Here $\beta_{1} = 1$}.
\end{figure}
%%%%%%%%%%%%%%%%%%%%%%%%%%%%%%%%%% %%%%%%%%%%%%%%%%%%%%%%%%%%%%%%%%%%%%%%%%%%%%%%%%%%%%%%%%%%
Again substituting (\ref{eq31}) in (\ref{eq30}), the scalar field is obtained as
\begin{equation}
\label{eq36}
\phi = \left[\frac{h(k + 2)}{2}\int{{(t^{n}e^{t})^{-\frac{3}{2}}}}dt\right]^{\frac{2}{(k + 2)}},
\end{equation}
Eq. (\ref{eq36}) implies that
\begin{equation}
\label{eq37}
\phi^{k} \dot{\phi}^{2} = h^{2} (t^{n}e^{t})^{-3}.
\end{equation}
Substituting (\ref{eq33})-(\ref{eq37}) in (\ref{eq11}) and (\ref{eq12}), the expressions for thermodynamical pressure ($p$) 
and energy density ($\rho$) for the model are obtained as
\begin{equation}
\label{eq38}
p = \frac{n}{t^{2}} - \frac{3}{4}\left(\frac{n}{t} + 1\right)^{2} + \left(\frac{1}{2}\omega h^{2} - \beta_{3}\right)(t^{n}e^{t})^{-3},
\end{equation}
\begin{equation}
\label{eq39}
\rho =  \frac{3}{4}\left(\frac{n}{t} + 1\right)^{2} + \left(\frac{1}{2}\omega h^{2} + \beta_{2}\right)(t^{n}e^{t})^{-3},
\end{equation}
where
\[
 \beta_{1} = b_{1}^{2} + b_{2}^{2} + b_{3}^{2},
\]
\[
 \beta_{2} = b_{1}b_{2} +  b_{2}b_{3} +  b_{3}b_{1},
\]
\begin{equation}
\label{eq40}
\beta_{3} = b_{2}^{2} + b_{3}^{2} + b_{2}b_{3}.
\end{equation}
In view of (\ref{eq29}), it is observed that the solutions given by (\ref{eq33})-(\ref{eq40}) satisfy the energy 
conservation equation (\ref{eq14}) identically and hence represent exact solutions of the Einstein's modified 
field equations (\ref{eq9})-(\ref{eq13}). From Eq. (\ref{eq39}), it is observed that $\rho$ is a 
positive decreasing function of time and it approaches to zero as $t \to \infty$. This behaviour is clearly 
depicted in Figure $2$ as a representative case with appropriate choice of constants of integrations 
and other physical parameters using reasonably well known situations. \\

The rate of expansion $H_{i}$ in the direction of $x$, $y$, and $z$ read as
\begin{equation}
\label{eq41}
H_{x} = \frac{1}{2}\left(\frac{n}{t} + 1\right) + b_{1}(t^{n}e^{t})^{-\frac{3}{2}},
\end{equation}
\begin{equation}
\label{eq42}
H_{y} = \frac{1}{2}\left(\frac{n}{t} + 1\right) + b_{2}(t^{n}e^{t})^{-\frac{3}{2}},
\end{equation}
\begin{equation}
\label{eq43}
H_{z} = \frac{1}{2}\left(\frac{n}{t} + 1\right) + b_{3}(t^{n}e^{t})^{-\frac{3}{2}}.
\end{equation}
The Hubble parameter, expansion scalar and shear of the model are, respectively given by
\begin{equation}
\label{eq44}
H = \frac{1}{2}\left(\frac{n}{t} + 1\right),
\end{equation}
\begin{equation}
\label{eq45}
\theta = \frac{3}{2}\left(\frac{n}{t} + 1\right),
\end{equation}
\begin{equation}
\label{eq46}
\sigma^{2} = \frac{1}{2}\beta_{1}(t^{n}e^{t})^{-3},
\end{equation}
where $\beta_{1}$ is given in Eq. (\ref{eq40}). \\

The spatial volume ($V)$ and anisotropy parameter ($A_{m}$) are found to be
\begin{equation}
\label{eq47}
V = (t^{n}e^{t})^{\frac{3}{2}},
\end{equation}
\begin{equation}
\label{eq48}
A_{m} = \frac{4}{3}\beta_{1}(t^{n}e^{t})^{-3}\left(\frac{n}{t} + 1\right)^{-2}.
\end{equation}
From Eqs. (\ref{eq47}) and (\ref{eq45}) we observe that the spatial volume is zero at $t = 0$ and the expansion scalar 
is infinite, which show that the universe starts evolving with zero volume at $t = 0$ which is big bang scenario. From 
Eqs. (\ref{eq33})-(\ref{eq35}), we observe that the spatial scale factors are zero at the initial epoch $t = 0$ and hence 
the model has a point type singularity \cite{ref46}). We observe that proper volume increases with time. \\

The dynamics of the mean anisotropic parameter depends on the constant $\beta_{1} = b_{1}^{2} + b_{2}^{2} + b_{3}^{2} $. 
From Eq. (\ref{eq48}), we observe that at late time when $t \to \infty$, $A_{m} \to 0$. Thus, our model has transition 
from initial anisotropy to isotropy at present epoch which is in good harmony with current observations. Figure $3$ depicts 
the variation of anisotropic parameter ($A_{m}$) versus cosmic time $t$. From the figure, we observe that $A_{m}$ decreases 
with time and tends to zero as $t \to \infty$. Thus, the observed isotropy of the universe can be achieved in our model at 
present epoch. \\

It is important to note here that $\lim_{t \to 0}\left(\frac{\rho}{\theta^{2}}\right)$ spread out to be constant. Therefore the 
model of the universe goes up homogeneity and matter is dynamically negligible near the origin. This is in good agreement with 
the result already given by Collins \cite{ref44}. \\ 
 
%%%%%%%%%%%%%%%%%%%%%%%%%%%%%%%%%%%%%%%%%%%%%%%%%%%%%%%%%%%%%%%%%%%%%%%%%%%%%%%%%%%%%%%%%%%%%%%%%%%%%%%%%%%%%%%%%
%%%%%%%%%%%%%%%%%%%%%%%%%%%%%%%%%%%%%%%%%% SUBSECTION 3.2 %%%%%%%%%%%%%%%%%%%%%%%%%%%%%%%%%%%%%%%%%%%%%%%%%%%%%%%%%%%%%%
\subsection{Case 2: when $a(t) = \sqrt{te^{t}}$}
Following Amirhashchi et al. \cite{ref42}, we consider the following {\it ansatz} for the scale factor, where 
the increase in terms of time evolution is 
\begin{equation}
\label{eq49}
a(t) = \sqrt{t e^{t}}.
\end{equation}
By the above choice of scale factor yields a time dependent deceleration parameter. Using (\ref{eq49}) into (\ref{eq18}), 
we find
\begin{equation}
\label{eq50} q = \frac{2}{(1 + t)^{2}} - 1.
\end{equation}
Using Eq. (\ref{eq49}) in (\ref{eq33})-(\ref{eq35}), we get the following expressions for metric coefficients
\begin{equation}
\label{eq51}
A = a_{1} (\sqrt{t e^{t}}) \exp{\left[b_{1}\int{(t e^{t})^{-\frac{3}{2}}}dt\right]},
\end{equation}
\begin{equation}
\label{eq52}
B = a_{2} (\sqrt{t e^{t}})\exp{\left[b_{2}\int{(t e^{t})^{-\frac{3}{2}}}dt\right]},
\end{equation}
\begin{equation}
\label{eq53}
C = a_{3} (\sqrt{t e^{t}})\exp{\left[b_{3}\int{(t e^{t})^{-\frac{3}{2}}}dt\right]},
\end{equation}
where the constants are already defined in previous section. \\

%%%%%%%%%%%%%%%%%%% Figure 4 %%%%%%%%%%%%%%%%%%%%%%%%%%%%%%%%%%%%%%%%%%%%%%%%%%%%%%%%%%%%%
\begin{figure}[ht]
\centering
\includegraphics[width=10cm,height=10cm,angle=0]{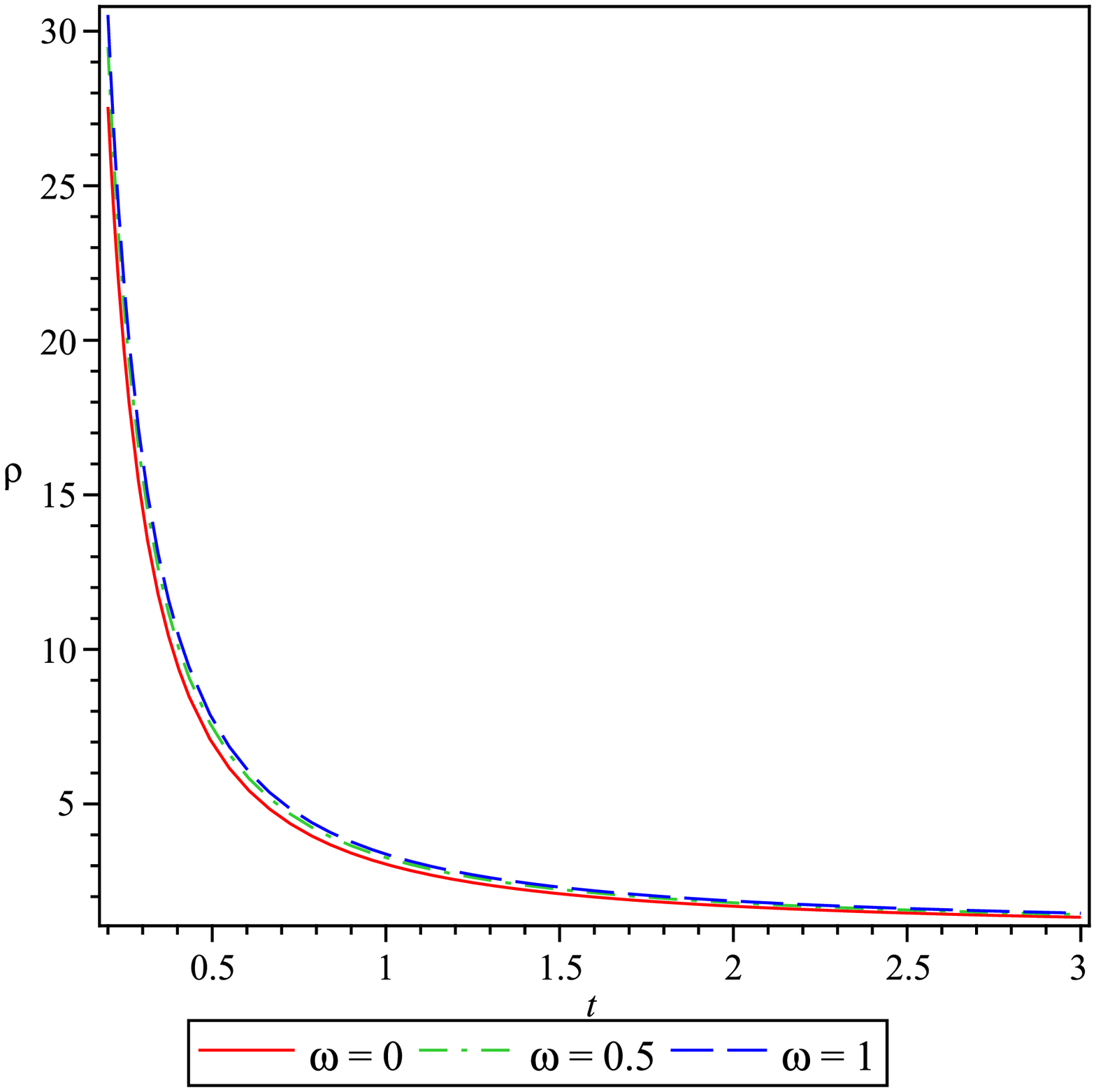} \\
\caption{The plot of energy density $\rho$ versus $t$.
Here $h= \beta_{2} = 1$}.
\end{figure}
%%%%%%%%%%%%%%%%%%%%%%%%%%%%%%%%%% %%%%%%%%%%%%%%%%%%%%%%%%%%%%%%%%%%%%%%%%%%%%%%%%%%%%%%%%%%

Again substituting (\ref{eq49}) in (\ref{eq30}), the scalar field is obtained as
\begin{equation}
\label{eq54}
\phi = \left[\frac{h(k + 2)}{2}\int{{(te^{t})^{-\frac{3}{2}}}}dt\right]^{\frac{2}{(k + 2)}},
\end{equation}
Eq. (\ref{eq54}) implies that
\begin{equation}
\label{eq55}
\phi^{k} \dot{\phi}^{2} = h^{2} (t e^{t})^{-3}.
\end{equation}
Substituting (\ref{eq51})-(\ref{eq55}) in (\ref{eq11}) and (\ref{eq12}), the expressions for thermodynamical pressure ($p$) 
and energy density ($\rho$) for the model are obtained as
\begin{equation}
\label{eq56}
p = \frac{1}{t^{2}} - \frac{3}{4}\left(\frac{1}{t} + 1\right)^{2} + \left(\frac{1}{2}\omega h^{2} - \beta_{3}\right)(t e^{t})^{-3},
\end{equation}
\begin{equation}
\label{eq57}
\rho =  \frac{3}{4}\left(\frac{1}{t} + 1\right)^{2} + \left(\frac{1}{2}\omega h^{2} + \beta_{2}\right)(t e^{t})^{-3},
\end{equation}
where $\beta_{2}$ and $\beta_{3}$ are already given in previous section. \\

%%%%%%%%%%%%%%%%%%% Figure 5 %%%%%%%%%%%%%%%%%%%%%%%%%%%%%%%%%%%%%%%%%%%%%%%%%%%%%%%%%%%%%
\begin{figure}[ht]
\centering
\includegraphics[width=10cm,height=10cm,angle=0]{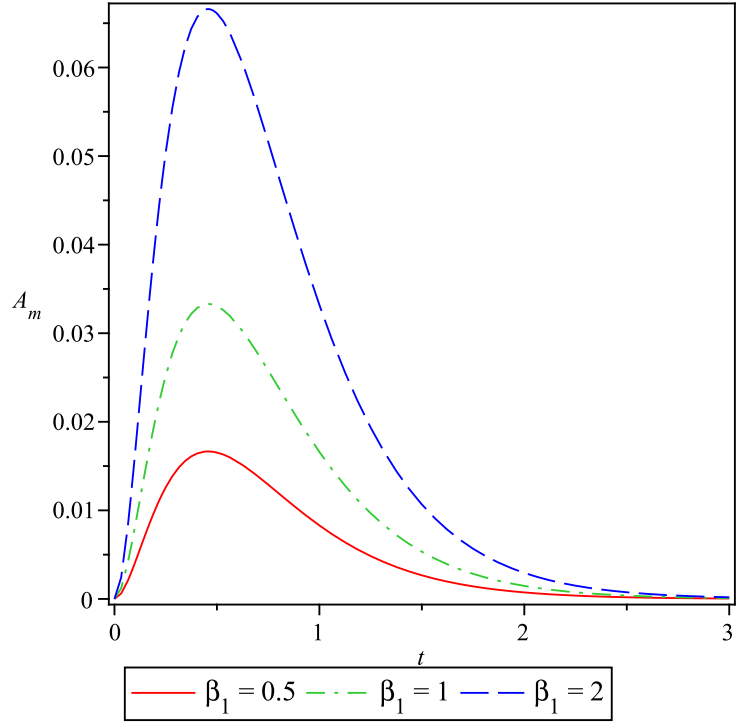} \\
\caption{The plot of anisotropic parameter $A_{m}$ versus $t$}.
\end{figure}
%%%%%%%%%%%%%%%%%%%%%%%%%%%%%%%%%% %%%%%%%%%%%%%%%%%%%%%%%%%%%%%%%%%%%%%%%%%%%%%%%%%%%%%%%%%%

In view of (\ref{eq29}), it is observed that the solutions given by (\ref{eq51})-(\ref{eq57}) satisfy the energy 
conservation equation (\ref{eq14}) identically and hence represent exact solutions of the Einstein's modified 
field equations (\ref{eq9})-(\ref{eq13}).  From Eq. (\ref{eq57}), it is observed that $\rho$ is a positive decreasing 
function of time and it approaches to zero as $t \to \infty$. Figure $4$ shows this behaviour of $\rho$.\\

The rate of expansion $H_{i}$ in the direction of $x$, $y$, and $z$ read as
\begin{equation}
\label{eq58}
H_{x} = \frac{1}{2}\left(\frac{1}{t} + 1\right) + b_{1}(t e^{t})^{-\frac{3}{2}},
\end{equation}
\begin{equation}
\label{eq59}
H_{y} = \frac{1}{2}\left(\frac{1}{t} + 1\right) + b_{2}(t e^{t})^{-\frac{3}{2}},
\end{equation}
\begin{equation}
\label{eq60}
H_{z} = \frac{1}{2}\left(\frac{1}{t} + 1\right) + b_{3}(t e^{t})^{-\frac{3}{2}}.
\end{equation}
The Hubble parameter, expansion scalar and shear of the model are, respectively given by
\begin{equation}
\label{eq61}
H = \frac{1}{2}\left(\frac{1}{t} + 1\right),
\end{equation}
\begin{equation}
\label{eq62}
\theta = \frac{3}{2}\left(\frac{1}{t} + 1\right),
\end{equation}
\begin{equation}
\label{eq63}
\sigma^{2} = \frac{1}{2}\beta_{1}(t e^{t})^{-3},
\end{equation}
The spatial volume ($V)$ and anisotropy parameter ($A_{m}$) are found to be
\begin{equation}
\label{eq64}
V = (t e^{t})^{\frac{3}{2}},
\end{equation}
\begin{equation}
\label{eq65}
A_{m} = \frac{4}{3}\beta_{1}(t e^{t})^{-3}\left(\frac{1}{t} + 1\right)^{-2}.
\end{equation}
where $\beta_{1}$ is already defined in previous section. \\

This model has the same properties as described in Case$1$. This is a particular case of previous one. 

%%%%%%%%%%%%%%%%%%%%%%%%%%%%%%%%%%%%%%%%%%%%%%%%%%%%%%%%%%%%%%%%%%%%%%%%%%%%%%%%%%%%%%%%%%%%%%%%%%%%%%%%%%%%%%%%%
%%%%%%%%%%%%%%%%%%%%%%%%%%%%%%%%%%%%%%%%%% SUBSECTION 3.2 %%%%%%%%%%%%%%%%%%%%%%%%%%%%%%%%%%%%%%%%%%%%%%%%%%%%%%%
\subsection{Case 3: when $ a(t) = -\frac{1}{t} + t^{2}$}

%%%%%%%%%%%%%%%%%%% Figure 6 %%%%%%%%%%%%%%%%%%%%%%%%%%%%%%%%%%%%%%%%%%%%%%%%%%%%%%%%%%%%%
\begin{figure}[ht]
\centering
\includegraphics[width=10cm,height=10cm,angle=0]{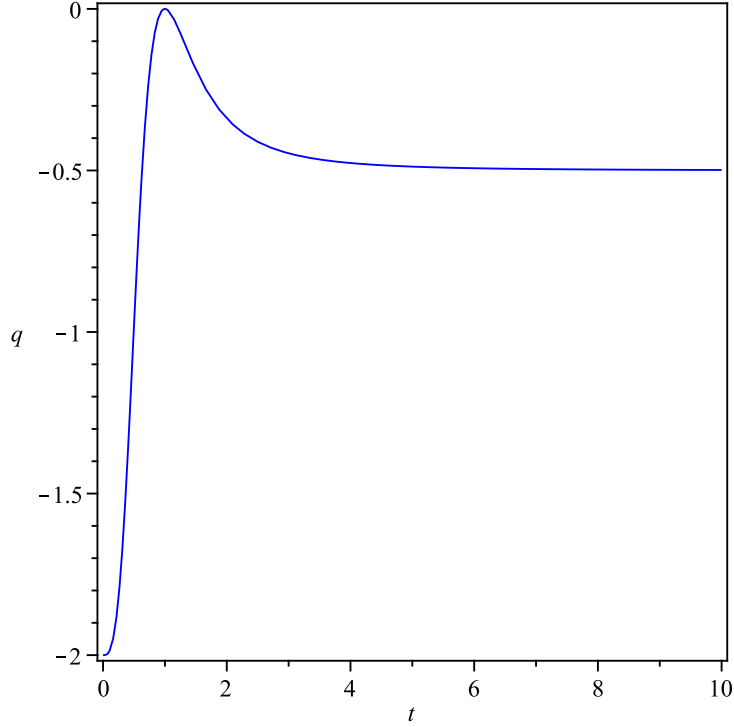} \\
\caption{The plot of deceleration parameter $q$ versus $t$}.
\end{figure}
%%%%%%%%%%%%%%%%%%%%%%%%%%%%%%%%%% %%%%%%%%%%%%%%%%%%%%%%%%%%%%%%%%%%%%%%%%%%%%%%%%%%%%%%%%%%

In this case, we consider the following {\it ansatz} for the scale factor, where 
the increase in terms of time evolution is 
\begin{equation}
\label{eq66}
a(t) = -\frac{1}{t} + t^{2},
\end{equation}
which yields a time dependent deceleration parameter. Using (\ref{eq66}) into (\ref{eq18}), 
we find
\begin{equation}
\label{eq67} q = -2\left(\frac{t^{3} - 1}{2t^{3} + 1}\right)^{2}.
\end{equation}

%%%%%%%%%%%%%%%%%%% Figure 7 %%%%%%%%%%%%%%%%%%%%%%%%%%%%%%%%%%%%%%%%%%%%%%%%%%%%%%%%%%%%%
\begin{figure}[ht]
\centering
\includegraphics[width=10cm,height=10cm,angle=0]{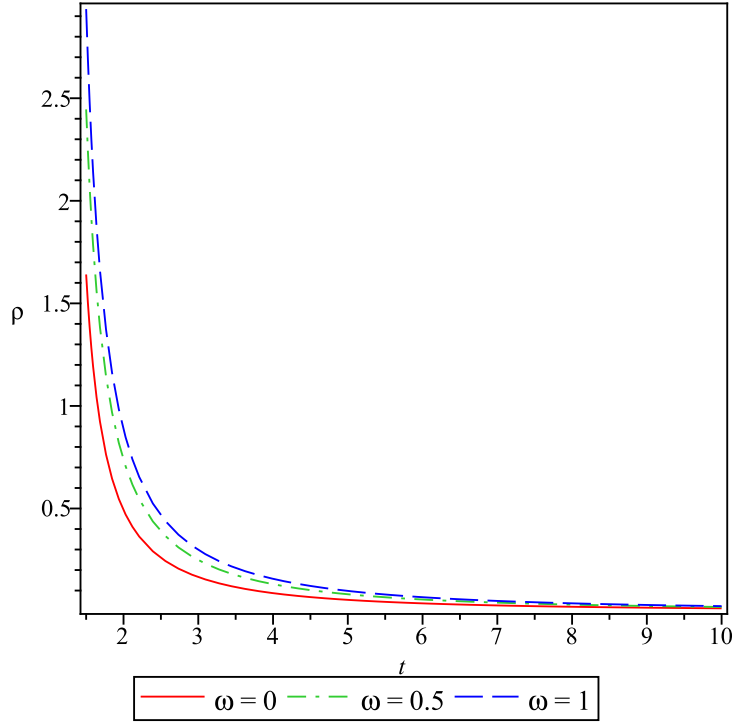} \\
\caption{The plot of energy density $\rho$ versus $t$.
Here $h= \beta_{2} = 1$}.
\end{figure}
%%%%%%%%%%%%%%%%%%%%%%%%%%%%%%%%%% %%%%%%%%%%%%%%%%%%%%%%%%%%%%%%%%%%%%%%%%%%%%%%%%%%%%%%%%%%

From Eq. (\ref{eq67}), we observe that $q = -2$ at $t = 0$ and $q = -0.5$ as $t \to \infty$. 
Figure $6$ depicts the variation of deceleration parameter versus cosmic time $t$. From figure 
we observe that $q = -2$ at $t = 0$ and increases very rapidly to reach its maximum near zero 
and then decreases with time and ultimately it becomes constant $(-0.5)$ as $t \to \infty$. 
The values of the deceleration parameter separate decelerating ($q > 0$) from accelerating ($q < 0$) 
periods in the evolution of the universe. Determination of the deceleration parameter from the count 
magnitude relation for galaxies is a difficult task due to the evolutionary effects. The present value 
$q_{0}$ of the deceleration parameter obtained from observations are $ -1.27 \leq q_{0} \leq 2$ 
(Schuecker et al. \cite{ref45}). Studies of galaxy counts from redshift surveys provide a value of
$q_{0} = 0.1$, with an upper limit of $q_{0} < 0.75$ \cite{ref45}. Recent observations show that the 
deceleration parameter of the universe is in the range $ - 1 \leq q \leq 0$ i.e $q_{0} \approx -0.77$ 
\cite{ref46}. Thus we see that the value of $q$ at present epoch in our derived model is very near to 
the value obtained by recent observation.\\ 
 
%%%%%%%%%%%%%%%%%%% Figure 8 %%%%%%%%%%%%%%%%%%%%%%%%%%%%%%%%%%%%%%%%%%%%%%%%%%%%%%%%%%%%%
\begin{figure}[ht]
\centering
\includegraphics[width=10cm,height=10cm,angle=0]{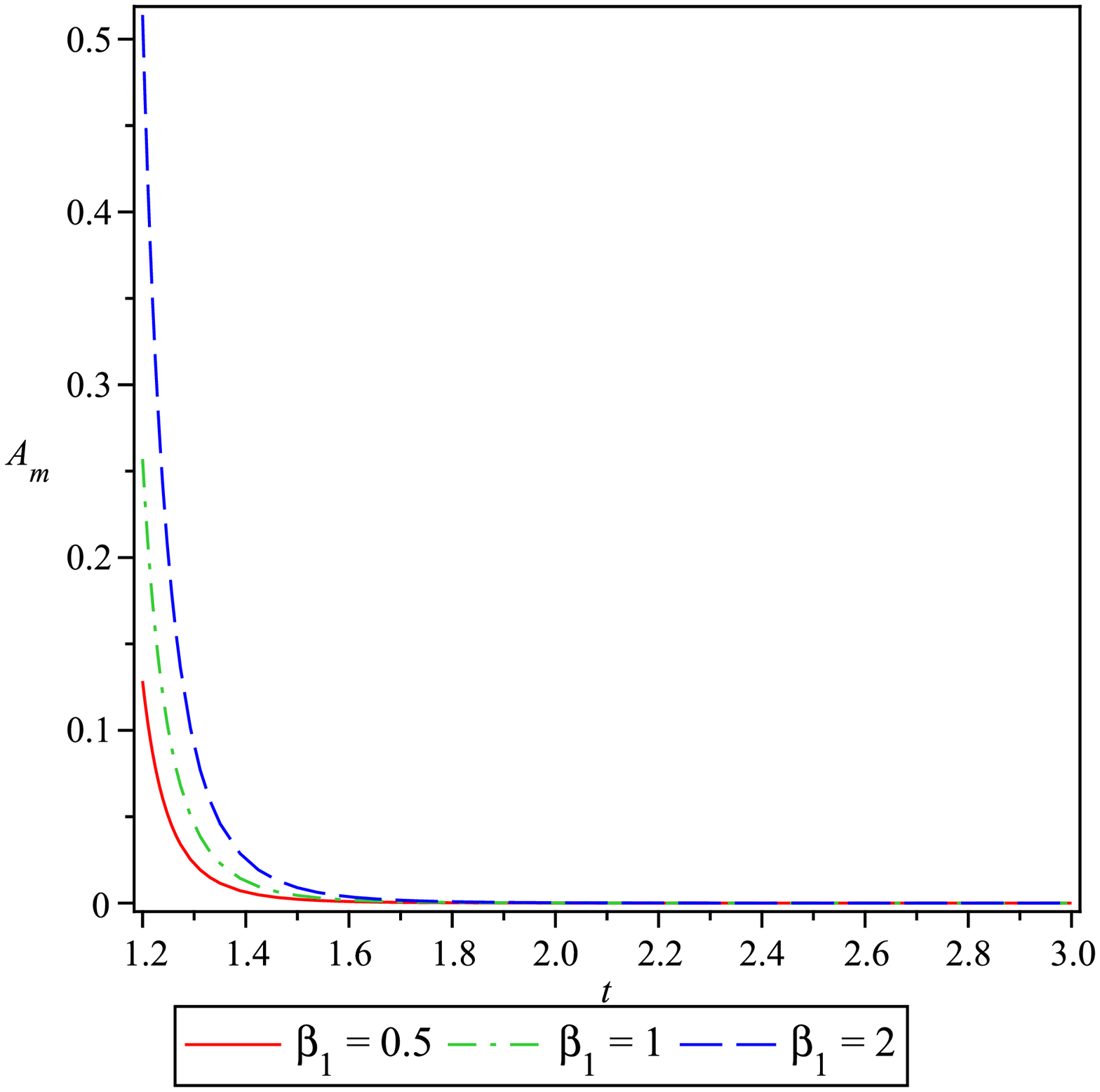} \\
\caption{The plot of anisotropic parameter $A_{m}$ versus $t$}.
\end{figure}
%%%%%%%%%%%%%%%%%%%%%%%%%%%%%%%%%% %%%%%%%%%%%%%%%%%%%%%%%%%%%%%%%%%%%%%%%%%%%%%%%%%%%%%%%%%%

Using Eq. (\ref{eq66}) in (\ref{eq33})-(\ref{eq35}), we get the following expressions for metric coefficients
\begin{equation}
\label{eq68}
A = a_{1} \left(-\frac{1}{t} + t^{2}\right) \exp{\Biggl[b_{1}\left(-\frac{1}{t} + t^{2}\right)dt\Biggr]},
\end{equation}
\begin{equation}
\label{eq69}
B = a_{2} \left(-\frac{1}{t} + t^{2}\right) \exp{\Biggl[b_{2}\left(-\frac{1}{t} + t^{2}\right)dt\Biggr]},
\end{equation}
\begin{equation}
\label{eq70}
C = a_{3} \left(-\frac{1}{t} + t^{2}\right) \exp{\Biggl[b_{3}\left(-\frac{1}{t} + t^{2}\right)dt\Biggl]},
\end{equation}
where the constants are already defined in previous section $3.1$. \\

Again substituting (\ref{eq66}) in (\ref{eq30}), the scalar field is obtained as
\begin{equation}
\label{eq71}
\phi = \left[\frac{h(k + 2)}{2}\int{{\left(-\frac{1}{t} + t^{2}\right)^{-3}}}dt\right]^{\frac{2}{(k + 2)}},
\end{equation}
Eq. (\ref{eq71}) implies that
\begin{equation}
\label{eq72}
\phi^{k} \dot{\phi}^{2} = h^{2}\left(-\frac{1}{t} + t^{2}\right)^{-6} .
\end{equation}
Substituting (\ref{eq68})-(\ref{eq72}) in (\ref{eq11}) and (\ref{eq12}), the expressions for isotropic 
pressure ($p$) and energy density ($\rho$) for the model are obtained as
\begin{equation}
\label{eq73}
p = \left(\frac{1}{2}\omega h^{2} - \beta_{3}\right)\left(-\frac{1}{t} + t^{2}\right)^{-6} - 
\frac{3(2t^{2} + 1)^{2}}{t^{2}(t^{3} - 1)^{2}},
\end{equation}
\begin{equation}
\label{eq74}
\rho = \left(\frac{1}{2}\omega h^{2} - \beta_{2}\right)\left(-\frac{1}{t} + t^{2}\right)^{-6} + 
\frac{3(2t^{2} + 1)^{2}}{t^{2}(t^{3} - 1)^{2}},
\end{equation}
where $\beta_{2}$ and $\beta_{3}$ are already given in section $3.1$. \\

In view of (\ref{eq29}), it is observed that the solutions given by (\ref{eq68})-(\ref{eq74}) 
satisfy the energy conservation equation (\ref{eq14}) identically and hence represent exact 
solutions of the Einstein's modified field equations (\ref{eq9})-(\ref{eq13}). From Eq. (\ref{eq74}), 
it is observed that $\rho$ is a positive decreasing function of time and it approaches to zero 
as $t \to \infty$. Figure $7$ depicts the variation of energy density versus cosmic time $t$. 
We observe from the figure that $\rho$ is a positive decreasing function of time and it approaches 
to zero at late time. \\

The rate of expansion $H_{i}$ in the direction of $x$, $y$, and $z$ read as
\begin{equation}
\label{eq75}
H_{x} = \frac{(2t^{3} + 1)}{(t^{3} - 1)t} + b_{1}\left(-\frac{1}{t} + t^{2}\right)^{-3},
\end{equation}
\begin{equation}
\label{eq76}
H_{y} = \frac{(2t^{3} + 1)}{(t^{3} - 1)t} + b_{2}\left(-\frac{1}{t} + t^{2}\right)^{-3},
\end{equation}
\begin{equation}
\label{eq77}
H_{z} = \frac{(2t^{3} + 1)}{(t^{3} - 1)t} + b_{3}\left(-\frac{1}{t} + t^{2}\right)^{-3}.
\end{equation}
The Hubble parameter, expansion scalar and shear of the model are, respectively given by
\begin{equation}
\label{eq78}
H = \frac{(2t^{3} + 1)}{(t^{3} - 1)t},
\end{equation}
\begin{equation}
\label{eq79}
\theta = \frac{3(2t^{3} + 1)}{(t^{3} - 1)t},
\end{equation}
\begin{equation}
\label{eq80}
\sigma^{2} = \frac{1}{2}\beta_{1}\left(-\frac{1}{t} + t^{2}\right)^{-6}
\end{equation}
The spatial volume ($V)$ and anisotropy parameter ($A_{m}$) are found to be
\begin{equation}
\label{eq81}
V = \left(-\frac{1}{t} + t^{2}\right)^{3}.
\end{equation}
\begin{equation}
\label{eq82}
A_{m} = \frac{1}{3}\beta_{1}\frac{(t^{3} - 1)^{2}t^{2}}{(2t^{3} + 1)^{2}} \left(-\frac{1}{t} + t^{2}\right)^{-6}. 
\end{equation}
where $\beta_{1}$ is already defined in previous section. \\

It is observed that the spatial volume is zero at $t = 1$ and the expansion scalar is infinite, which shows that 
the universe starts evolving with zero volume at initial epoch $t = 1$ with initial rate of expansion. The scale 
factors also vanish at $t = 1$ and hence the model has a Point type singularity \cite{ref47} at the initial epoch.
The pressure, energy density, Hubble's factors and shear scalar diverge at the initial singularity. Figure $8$ is 
the plot of the anisotropic parameter $(A_{m}$) versus cosmic time $t$. From the figure, we observe that $A_{m}$ 
decreases with time and tends to zero  as $t \to \infty$ for all values of $\beta_{1}$. Thus, the observed isotropy 
of the universe can be achieved in our derived model at present epoch. The shear tensor also tends to zero in this model.
Thus the model represents shearing, non-rotating and expanding model of the universe with a big bang start approaching 
to isotropy at late times.   \\

%%%%%%%%%%%%%%%%%%%%%%%%%%%%%%%  SECTION 4  %%%%%%%%%%%%%%%%%%%%%%%%%%%%%%%%%%%%%%%%%%%%%%%%%%%%%%
\section{Concluding Remarks}
In this paper we have studied a spatially homogeneous and anisotropic Bianchi type-I space-time within the 
framework of the scalar-tensor theory of gravitation proposed by S$\acute{a}$ez and Ballester \cite{ref3}. The field 
equations have been solved exactly with suitable physical assumption. The solutions in all three cases satisfy 
the energy conservation Eq. (\ref{eq14}) identically. Therefore, exact and physically viable Bianchi type-I models 
have been obtained. To find the deterministic solution, we have considered three types of scale factors which yield 
three different time dependent deceleration parameters. \\

In Case $1$, we choosed the scale factor $a(t) = \sqrt{t^{n}e^{t}}$ which yields a time dependent deceleration 
parameter $q = \frac{2n}{(n + t)^{2}} - 1$. Since all the scale factors vanish at initial moment, hence the 
model has a big bang singularity at $t = 0$. This is a Point type singularity. This case represents a 
model which generates a transition of the universe from early decelerated phase to the recent accelerating phase 
which is in good agreement with the recent observations (Caldwell et al. \cite{ref48}) (see, Fig. $1$) whereas in 
Pradhan et al. \cite{ref39} only the evolution takes place in an accelerating phase. The model has also transition from initial 
anisotropy to isotropy at late time (i.e. present epoch) which is in good harmony with current observations (see, Fig. $3$). For 
different choice of n, we can generate a class of cosmological models in Bianchi type-I space-time. It is observed 
that such models are also in good harmony with current observations [26, 27, 31-33].\\

In Case $2$, we considered the scale factor $a(t) = \sqrt{te^{t}}$ which yields a time dependent deceleration 
parameter $q = \frac{2}{(1 + t)^{2}} - 1$. This is a particular case of previous one. This model has the same 
properties as obtained in Case $1$. It is worth mention here that the anisotropic parameter becomes negligible 
at $t \to \infty$ which, in turn, implies that the derived model isotropizes at present epoch.  \\

In Case $3$, we considered the scale factor $a(t) = -\frac{1}{t} + t^{2}$ which yields a time dependent deceleration 
parameter $q = -2\left(\frac{t^{3} - 1}{2t^{3} + 1}\right)^{2}$. In this case we get an accelerating universe 
(see, Fig. $6$) at present epoch. In this case, the anisotropic behaviour of the universe dies out on later times and 
the observed isotropy of the universe can be achieved in derived model at present epoch (see, Fig. $8$). The model 
has a point type singularity at initial epoch $t = 1$. \\

It is worth mentioned here that in all above three cases, their anisotropic solutions may enter to some isotropic 
inflationary era. All the above three models represent expanding, shearing and non-rotating universe, which approach 
to isotropy for large value of $t$. This is consistent with the behaviour of the present universe as already discussed 
in introduction. If we set $k = 0$, then the solutions reduce to the solution in general relativity and the 
S$\acute{a}$ez-Ballester scalar-tensor theory of gravitation tends to standard general theory of relativity in every 
respect. In literature we can get the solutions of the field equations of S$\acute{a}$ez-Ballester scalar-tensor theory 
of gravitation by using a constant deceleration parameter. So the solutions presented in this paper are new and different 
from other author's solutions. \\

Finally, our solutions may be useful for better understanding of the evolution of the universe in Bianchi-I space-time within 
the framework of S$\acute{a}$ez-Ballester scalar-tensor theory of gravitation. The solutions presented here can be one of 
the potential candidates to describe the observed universe.      
%%%%%%%%%%%%%%%%%%%%%%%%%%%%%%%%%%%%%%%%%%%%%%%%%%%%%%%%%%%%%%%%%%%%%%%%%%%%%%%%%%%%%%%%%%%%%%%%%%%%%%%% 
\section*{Acknowledgement} 
Author (A. Pradhan) would like to thank the Inter-University Centre for Astronomy and Astrophysics (IUCAA), Pune, India 
for providing facility and support under associateship program where part of this work was carried out. The financial 
support (Project No. C.S.T./D-1536) in part by State Council of Science and Technology, Uttar Pradesh, India is gratefully 
acknowledged by A. Pradhan. 
%%%%%%%%%%%%%%%%%%%%%%%%%%%%%%%%%%%%%%%%%%%%%%%%%%%%%%%%%%%%%%%%%%%%%%%%%%%%%%%%%%%%%%%%%%%%%%%%%%%%%%%%%%

\end{document}